\documentclass{svproc}

\usepackage{graphicx}
\usepackage{float}               
\usepackage{amsmath,amssymb}
\usepackage{booktabs,longtable,array,multirow}
\usepackage{xcolor}
\usepackage[hidelinks]{hyperref}
\usepackage{xurl}
\usepackage[numbers]{natbib}
\usepackage[section]{placeins}
\usepackage{microtype}
\usepackage{subcaption}
\usepackage[T1]{fontenc}
\usepackage[utf8]{inputenc}
\usepackage{lmodern}
\usepackage[T1]{fontenc}
\usepackage{pdfpages}

\usepackage{etoolbox}
\newcommand{\gene}[1]{\textit{#1}}

\begin{document}
\pretocmd{\section}{\setcounter{enumiv}{0}}{}{}

\mainmatter

\title{Network Community Detection and Novelty Scoring Reveal Underexplored Hub Genes in Rheumatoid Arthritis}
\titlerunning{Network Community Detection in RA}

\author{Neda Amirirad\inst{1} \and Hiroki Sayama\inst{1,2,3}}
\authorrunning{Amirirad and Sayama}

\institute{
School of Systems Science and Industrial Engineering, Binghamton University, Binghamton, NY, USA
\and
Binghamton Center of Complex Systems, Binghamton University, Binghamton, NY, USA
\and
Waseda Innovation Lab, Waseda University, Tokyo, Japan\\
\email{neda.amirirad@binghamton.edu} \quad
\email{sayama@binghamton.edu}
}

\maketitle

\begin{abstract}
Understanding the modular structure and central elements of complex biological networks is critical for uncovering system-level mechanisms in disease. Here, we constructed weighted gene co-expression networks from bulk RNA-seq data of rheumatoid arthritis (RA) synovial tissue, using pairwise correlation and a percolation-guided thresholding strategy. Community detection with Louvain and Leiden algorithms revealed robust modules, and node-strength ranking identified the top 50 hub genes globally and within communities. To assess novelty, we integrated genome-wide association studies (GWAS) with literature-based evidence from PubMed, highlighting five high-centrality genes with little to no prior RA-specific association. Functional enrichment confirmed their roles in immune-related processes, including adaptive immune response and lymphocyte regulation. Notably, these hubs showed strong positive correlations with T- and B-cell markers and negative correlations with NK-cell markers, consistent with RA immunopathology. Overall, our framework demonstrates how correlation-based network construction, modularity-driven clustering, and centrality-guided novelty scoring can jointly reveal informative structure in omics-scale data. This generalizable approach offers a scalable path to gene prioritization in RA and other autoimmune conditions.
\end{abstract}
\keywords{Gene co-expression networks, Community detection, Rheumatoid arthritis, Hub genes, Novelty scoring}

\section{Introduction}

Rheumatoid arthritis (RA) is a chronic systemic autoimmune disease that leads to persistent synovial inflammation, progressive joint destruction, and functional disability \citep{firestein2009,smolen2016ra}. It affects approximately 18--23 million people worldwide \citep{gkatzionis2017raepidemiology}, imposing a substantial societal and economic burden due to reduced quality of life, loss of productivity, and increased healthcare costs \citep{choi2019economicRA}. Given its multifactorial etiology and heterogeneous clinical outcomes, RA requires systems-level approaches to integrate high-dimensional molecular data with network perspectives \citep{barabasi2011}. Traditional differential expression analyses, although widely used, often overlook genes that are functionally important yet not strongly differentially expressed \citep{robinson2010edgeR,love2014deseq2}. In contrast, analysis of gene co-expression patterns offers a complementary strategy by focusing on relationships between genes and the modular organization of the transcriptome \citep{zhang2005,wgcna}. We expect that applying community detection to RA transcriptomic data will uncover modular gene expression patterns and highlight novel hub genes that may contribute to disease heterogeneity and immune dysregulation.

Network-based approaches have become increasingly important for characterizing the modular structure of biological systems. Methods such as weighted gene co-expression network analysis (WGCNA) \citep{zhang2005,wgcna} can identify gene modules associated with disease phenotypes, while community detection algorithms, including Louvain \citep{blondel2008fast} and Leiden \citep{traag2019}, provide scalable strategies for uncovering robust network partitions. A key challenge in constructing correlation-based networks is the choice of threshold; conventional fixed cutoffs may either lose relevant biological edges or retain noise. To address this, percolation-based thresholding has been introduced as a principled approach to balance sparsity and connectivity in weighted complex networks \citep{esfahlani2017percolation}.

Beyond module detection, functional enrichment analysis provides critical biological interpretation. Curated resources such as Gene Ontology (GO) \citep{ashburner2000go,goconsortium2021}, KEGG \citep{kanehisa2000kegg}, Reactome \citep{reactome2016}, and WikiPathways \citep{slenter2018wikipathways}, together with computational platforms such as g:Profiler \citep{raimondogprofiler}, allow systematic identification of overrepresented processes. Furthermore, integrating GWAS repositories like the NHGRI-EBI Catalog \citep{buniello2019nhgri} and literature-based measures such as PubMed co-mentions \citep{sayers2020pubmed} supports the evaluation of novelty for candidate genes. This combined strategy has been applied to highlight uncharacterized drivers of autoimmune disease mechanisms, including RA and systemic lupus erythematosus (SLE) \citep{tyagi2023}.

Emerging applications underscore the potential of such approaches in specific RA contexts. For example, transcriptional modules have been linked to disease progression during pregnancy \citep{wright2023}, revealing dynamic regulation of immune pathways. Complementary tools such as CIBERSORT enable the estimation of immune infiltration directly from bulk transcriptomes \citep{newman2015cibersort}, while single-cell RNA-seq offers enhanced resolution into immune heterogeneity \citep{luecken2019scrna}.

Taken together, these advances demonstrate the power of network-based systems biology to uncover disease-relevant modules and candidate genes in RA. By integrating correlation-based network construction, community detection, enrichment analysis, and novelty assessment, it becomes possible to derive mechanistic insights into RA pathogenesis that extend beyond conventional differential expression analysis, potentially informing biomarker discovery and therapeutic strategies.

\section{Materials and Methods}

We analyzed bulk RNA-seq data from the Pathobiology of Early Arthritis Cohort (PEAC), which includes synovial tissue samples from $n=87$ rheumatoid arthritis (RA) patients \cite{lewis2019}. The original dataset contained 19,279 protein-coding genes with associated clinical annotations. To reduce noise and enhance interpretability, genes with low expression or broadly non-specific functions (e.g., mitochondrial or ribosomal genes) were removed \cite{luecken2019scrna}. We then retained 2,772 genes based on prior differential expression analyses in the PEAC study~\cite{lewis2019}, and used this panel consistently in all downstream network analyses.

To infer co-expression patterns, we computed the pairwise Pearson correlation between all gene pairs using the filtered expression matrix. Negative correlations were discarded, and diagonal entries were zeroed to exclude self-similarity. The resulting $2772 \times 2772$ non-negative correlation matrix was treated as a weighted adjacency matrix for network construction.

To convert the weighted gene co-expression network into an unweighted graph while preserving its global topology, we used the percolation-based thresholding method proposed in~\cite{esfahlani2017percolation}. Let $n(\theta)$ denote the size of the largest connected component (LCC) after thresholding at correlation $\theta$, and let $n_0$ be the LCC size in the original weighted network (typically equal to the number of nodes). This method scans $\theta$ from high to low and identifies the largest threshold satisfying $n(\theta_c) = \alpha n_0$, where $\alpha$ is an adjustable parameter ($\alpha = 1$ in the original study~\cite{esfahlani2017percolation}).

In our implementation, we scanned $\theta \in [0.30, 0.80]$ with a step of 0.02, retained only positive correlations, set the diagonal to zero, and constructed an undirected graph at each $\theta$. For each graph we recorded: (i) fraction of nodes in the LCC $n(\theta)/n_0$, (ii) number of connected components, and (iii) edge count.

Hub genes were identified based on node strength, defined as the sum of edge weights connected to each node. We ranked all genes by strength and selected the top 50 as global hubs. Additionally, to capture community-specific centrality, we identified the top 20 hubs per community based on intra-community strength, applying a minimum community size filter to avoid unstable results. Duplicates were resolved, and final lists were re-ranked globally.

To assess the novelty of candidate hub genes, we implemented a two-step screening pipeline. First, each gene was queried in the NHGRI-EBI GWAS Catalog~\cite{buniello2019nhgri} for known genome-wide significant associations with RA. Second, we queried PubMed \cite{sayers2020pubmed} for literature co-mentions using the phrase ``\texttt{<GENE> AND rheumatoid arthritis}''. Genes were then classified into three novelty tiers: \textit{high novelty} (no GWAS associations and zero PubMed hits), \textit{medium novelty} (no GWAS, and $\leq$ 3 PubMed hits), and \textit{known} (GWAS-associated or $>$3 co-mentions).

Functional enrichment analysis was conducted using the g:Profiler tool~\cite{raimondogprofiler}, targeting GO terms (BP, MF, CC; \cite{ashburner2000go,goconsortium2021}), Reactome~\cite{reactome2016}, KEGG~\cite{kanehisa2000kegg}, and WikiPathways~\cite{slenter2018wikipathways} databases. We focused on the five high-novelty hub genes and used the default human genome (Ensembl) background. A relaxed significance threshold (FDR-adjusted $p < 0.05$) was applied to ensure broader biological interpretability.

To investigate immune relevance, we evaluated the correlation between each high-novelty gene and predefined immune cell marker panels representing T, B, and NK cells \cite{bindea2013immunome,newman2015cibersort}. Median expression scores of each panel were computed across samples, and Spearman correlation coefficients were calculated between gene expression and panel medians, as well as individual markers.

Finally, we assessed the robustness of our results under varying network thresholds. By computing the Jaccard similarity between global Top-50 hub gene sets across different thresholds, we evaluated the stability of gene rankings and the reproducibility of key findings. All analyses were conducted in Python using open-source packages including \texttt{pandas}, \texttt{numpy}, \texttt{networkx}, \texttt{python-louvain}, and \texttt{leidenalg}. The entire pipeline is available upon request for reproducibility.

\section{Results}

\begin{figure}[ht]
  \centering  \includegraphics[width=0.95\linewidth]{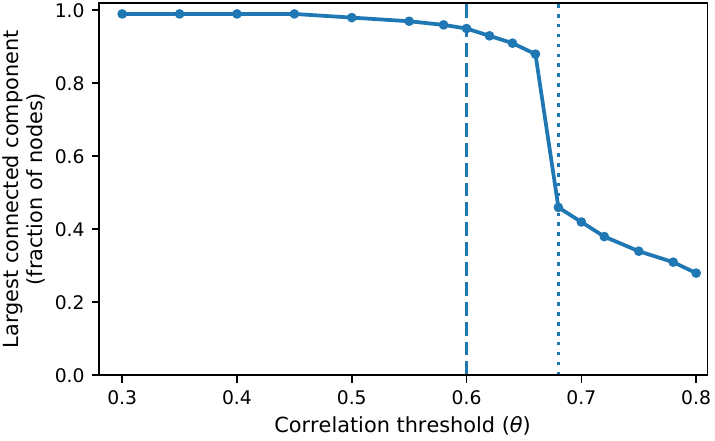}
  \caption{Percolation sweep: fraction of nodes in the largest connected component (LCC) 
  as a function of the correlation threshold $\theta$. 
  The dotted vertical line marks the formal percolation threshold at $\theta \approx 0.68$, 
  where the LCC shows an abrupt drop. 
  The dashed vertical line marks the chosen analysis threshold $\theta = 0.60$, 
  selected as a balance between global connectivity and modularity for downstream analyses.}
  \label{fig:perc_lcc}
\end{figure}

Figure~\ref{fig:perc_lcc} shows the percolation sweep of the largest connected component (LCC) as a function of the correlation threshold $\theta$. A pronounced transition (largest successive drop in LCC fraction, $\approx 0.383$) was observed at $\theta \approx 0.68$, indicating a sharp fragmentation of the network; we therefore consider $\theta \approx 0.68$ the formal percolation threshold in a statistical physics sense.

For the percolation threshold method, we adopted a relaxed criterion of $\alpha = 0.9$, requiring 90\% of nodes to remain in the LCC, because many genes in our dataset may be peripheral or less informative, corresponding to $\theta = 0.60$. This value was selected as the analysis threshold for downstream community detection and robustness checks. Accordingly, both $\theta \approx 0.68$ (true percolation threshold) and $\theta = 0.60$ (analysis threshold) are reported to transparently characterize the network’s percolation behavior.

\FloatBarrier

\begin{figure}[ht]
  \centering
\includegraphics[width=0.9\textwidth]{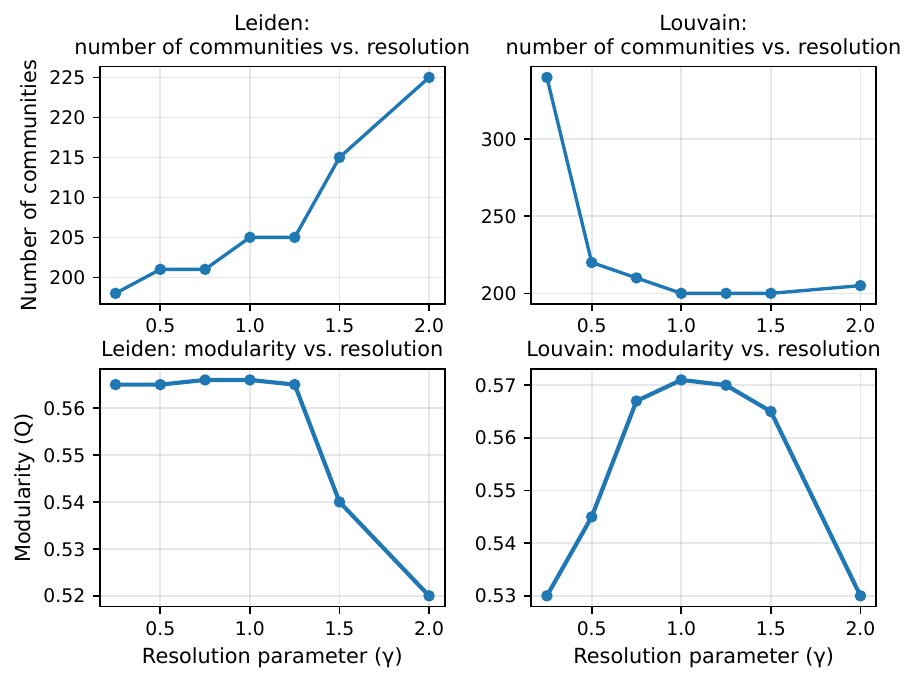}
  \caption{Resolution sweeps for Leiden (left) and Louvain (right). 
  Top panels: number of detected communities as a function of resolution parameter $\gamma$. 
  Bottom panels: modularity ($Q$) as a function of $\gamma$. 
  Leiden exhibited a stable plateau between $\gamma = 0.9$ and $1.2$, while Louvain peaked at $\gamma = 1.0$. 
  In both methods, modularity remained above $Q > 0.56$, confirming robust community structure.}
  \label{fig:resolution_sweeps}
\end{figure}

We then performed resolution sweeps to fine-tune community detection. 
As shown in Figure~\ref{fig:resolution_sweeps}, Leiden exhibited a plateau of high modularity 
between $\gamma = 0.9$ and $1.2$, while Louvain peaked at $\gamma = 1.0$. 
Both methods consistently produced modular networks with $Q > 0.56$, 
and the number of detected communities followed heavy-tailed distributions across resolutions (plots not shown). 
We therefore selected $\gamma = 1.0$ for both methods in downstream analyses. 
All runs used a fixed random seed for reproducibility.

Next, we identified hub genes using the node strength measure. The global top-50 hubs were extracted for both Louvain and Leiden networks. A notable overlap was observed between the two rankings, indicating robustness of centrality signals across methods. Table~\ref{tab:top50_hubs} shows a representative subset of shared hub genes common to both Louvain and Leiden rankings. A total of 40 such genes were identified, and 5 are shown for illustration.

\begin{table}[ht]
\centering
\caption{Representative subset of shared hub genes based on node strength (degree-weighted) in both Louvain and Leiden networks. Full list includes 40 genes with overlap across both partitions.}
\label{tab:top50_hubs}
\begin{tabular}{lcc}
\toprule
Gene & Strength (Louvain) & Strength (Leiden) \\
\midrule
\gene{SASH3} & 0.2357 & 0.2357 \\
\gene{SP140} & 0.2357 & 0.2357 \\
\gene{IL21R} & 0.2360 & 0.2360 \\
\gene{MYBL2} & 0.2342 & 0.2342 \\
\gene{SLAMF1} & 0.2339 & 0.2339 \\
\ldots & \ldots & \ldots \\
\bottomrule
\end{tabular}
\end{table}
\FloatBarrier

To assess the novelty of key genes, we queried six candidates against the GWAS Catalog and PubMed. As summarized in Table~\ref{tab:novelty_results}, five of the six showed no prior association with RA in GWAS and had zero PubMed co-mentions, qualifying them as high-novelty.

\begin{table}[ht]
\centering
\caption{Novelty classification of candidate genes based on GWAS and PubMed queries.}
\label{tab:novelty_results}
\begin{tabular}{lccc}
\toprule
Gene & GWAS-RA & PubMed-RA Count & Novelty \\
\midrule
\gene{P2RY8} & No & 0 & High \\
\gene{SASH3} & No & 0 & High \\
\gene{SIT1} & No & 0 & High \\
\gene{SNX20} & No & 0 & High \\
\gene{SP140} & No & 0 & High \\
\gene{NUP210} & No & 4 & Known \\
\bottomrule
\end{tabular}
\end{table}
\FloatBarrier

\begin{figure}[ht]
\centering
\includegraphics[width=0.75\textwidth]{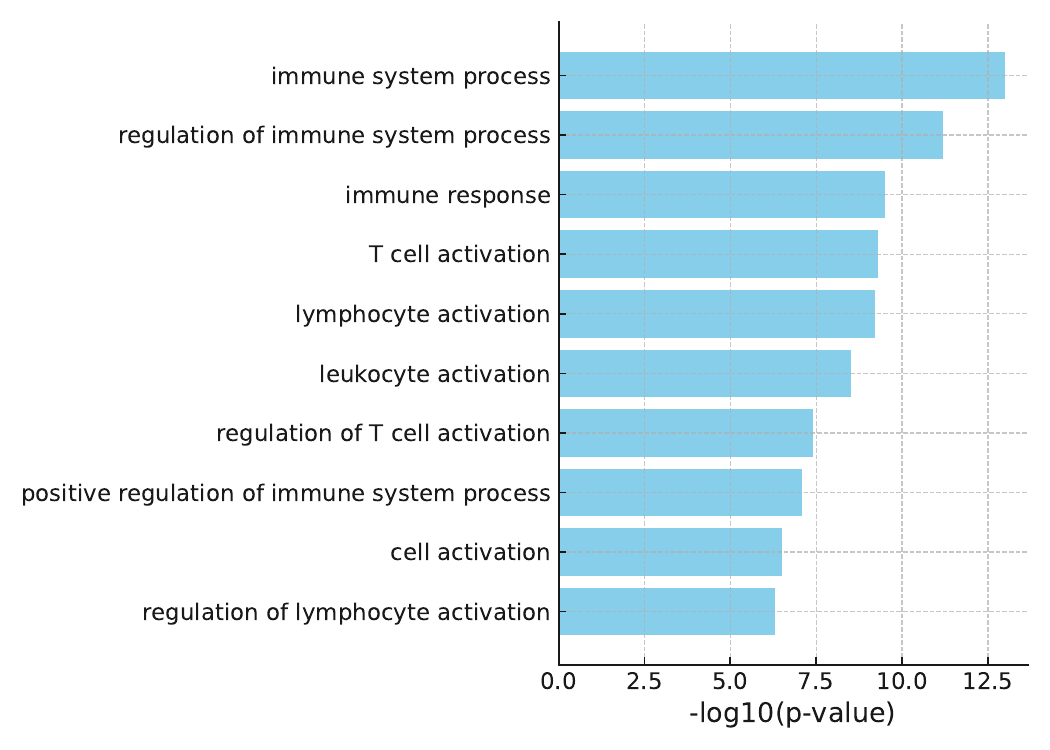}
\caption{Top enriched GO:BP terms for the five high-novelty genes using g:Profiler.}
\label{fig:enrichment_plot}
\end{figure}

We then performed functional enrichment of these genes using g:Profiler \cite{raimondogprofiler} against GO, Reactome, KEGG, and WikiPathways \cite{ashburner2000go,goconsortium2021,reactome2016,kanehisa2000kegg,slenter2018wikipathways}. Despite the small set, several immune-related terms were enriched, including ``lymphocyte homeostasis'' and ``adaptive immune response'' \cite{janeway2012} (Figure~\ref{fig:enrichment_plot}).

\FloatBarrier

\begin{figure}[ht]
\centering
\includegraphics[width=0.9\textwidth]{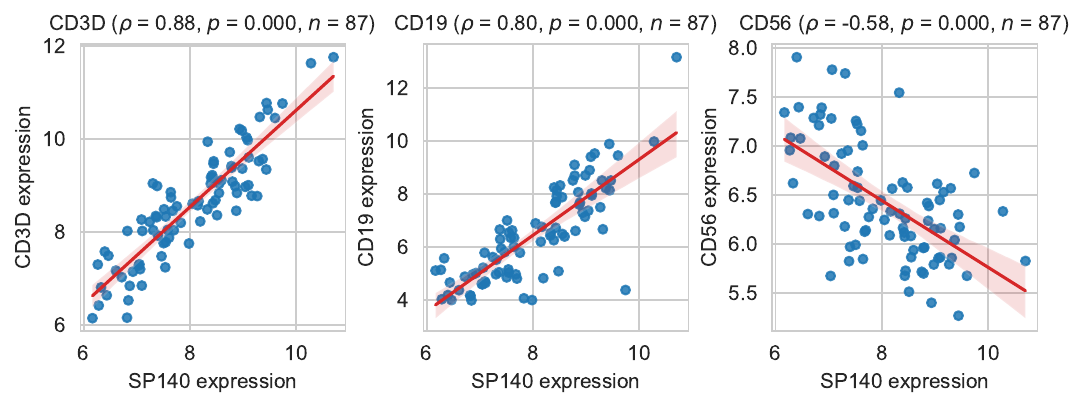}
\caption{Spearman correlation of \gene{SP140} expression with immune cell markers (\gene{CD3D}, \gene{CD19}, and \gene{CD56}). \gene{SP140} expression was strongly correlated with \gene{CD3D} and \gene{CD19} ($\rho = 0.88$ and $\rho = 0.80$, respectively; $p < 0.001$), and negatively correlated with \gene{CD56} ($\rho = -0.58$, $p < 0.001$).}
\label{fig:sp140_corr}
\end{figure}

To evaluate biological relevance, we correlated each novel hub gene with established immune cell markers. All five high-novelty genes (\gene{SASH3}, \gene{SP140}, \gene{SNX20}, \gene{SIT1}, and \gene{P2RY8}) showed strong positive correlations with T-cell (\gene{CD3D}) and B-cell (\gene{CD19}) markers, while consistently displaying strong negative correlations with the NK-cell marker (\gene{CD56}) (all $p < 0.001$). These results suggest that the identified hub genes are closely aligned with adaptive immune activity in RA, while inversely associated with NK cell signatures. Figure~\ref{fig:sp140_corr} illustrates representative correlations for \gene{SP140} as an example.

\FloatBarrier

\begin{figure}[ht]
\centering
\includegraphics[width=0.5\textwidth]{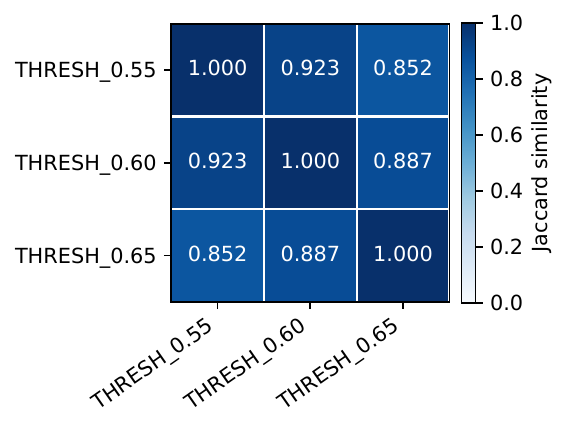}
\caption{Jaccard similarity between top-50 hub gene sets identified at thresholds $\theta = 0.55$, $\theta = 0.60$, and $\theta = 0.65$. High similarity values ($0.85$--$0.92$) indicate that hub rankings were robust and stable across adjacent thresholds.}
\label{fig:jaccard_heatmap}
\end{figure}

Finally, we assessed the sensitivity of our findings to correlation thresholding by comparing the top-50 hub gene sets obtained at $\theta = 0.55$, $\theta = 0.60$, and $\theta = 0.65$. Pairwise Jaccard similarity scores (0.85--0.92) indicated substantial overlap, and the resulting heatmap (Figure~\ref{fig:jaccard_heatmap}) confirms that hub rankings were relatively stable across adjacent thresholds.

\FloatBarrier

Overall, the combination of graph analysis, novelty mining, and enrichment revealed potentially underexplored regulators in RA pathogenesis, with robust support from network topology and immune relevance.

\section{Discussion}

In this study, we constructed a gene co-expression network from synovial RNA-seq data in the PEAC RA cohort and applied modularity-based community detection to uncover disease-relevant gene clusters. We further integrated hub gene analysis, novelty assessment, and functional enrichment to identify candidate regulators potentially involved in RA pathogenesis.

A major strength of our approach lies in the robustness of key hub genes across varying network construction strategies. Sensitivity analyses revealed that genes such as \gene{SASH3}, \gene{SP140}, and \gene{SNX20} consistently ranked among the top hubs across different thresholds and edge selection criteria. The Jaccard similarity analysis between top-50 hub gene sets confirmed the stability of these findings, reinforcing confidence in their biological relevance. Furthermore, the consistent detection of these hubs in both Louvain and Leiden partitions highlights their centrality irrespective of algorithmic variation.

To prioritize biologically meaningful yet potentially overlooked genes, we implemented a dual-screen novelty assessment combining the absence of GWAS associations and low PubMed co-mention counts \cite{sayers2020pubmed}. This process identified five high-novelty genes (\gene{P2RY8}, \gene{SASH3}, \gene{SIT1}, \gene{SNX20}, and \gene{SP140}) with little or no prior linkage to RA. This illustrates the utility of integrating topological centrality with novelty metrics to surface previously underexplored candidates.

Despite the limited gene set, functional enrichment of the high-novelty genes revealed convergence on immune-related biological processes such as lymphocyte homeostasis and adaptive immune response \cite{janeway2012}, which are highly relevant to RA pathophysiology \cite{smolen2016ra}. This functional coherence supports the hypothesis that these genes may play synergistic roles within immune regulatory networks.

We also evaluated the immune relevance of these genes by correlating their expression profiles with established markers of T cells, B cells, and NK cells \cite{janeway2012,bindea2013immunome,newman2015cibersort}. Notably, \gene{SASH3} and \gene{SP140} exhibited strong associations with T- and B-cell markers (\gene{CD3D}, \gene{CD19}), and negative correlations with the NK-cell marker \gene{CD56}, consistent with the predominance of adaptive immune activity in RA synovium.

Importantly, the robustness of our results across multiple thresholds and detection algorithms strengthens the credibility of our findings. In particular, adopting a relaxed percolation criterion ($\alpha=0.9$) justified the use of $\theta = 0.60$, which provided a favorable balance between connectivity and modularity while retaining biologically coherent results. Enrichment analyses further confirmed that the identified functions were specific to the RA transcriptomic landscape, rather than general trends.

Several limitations must be acknowledged. Co-expression networks reflect correlation, not causation, and may capture indirect relationships. Our novelty filter relies on existing curated databases, which may not reflect the most recent discoveries. Moreover, the use of bulk RNA-seq data may obscure cell-type-specific expression patterns. Future studies leveraging single-cell transcriptomics, spatial profiling, or perturbation-based experiments will be essential to validate and refine these findings.

Taken together, our integrative framework---combining network topology, novelty filtering, and immune validation---offers a scalable and interpretable strategy for gene prioritization in complex immune-mediated diseases. These findings underscore the power of network-based prioritization in surfacing underexplored yet biologically coherent candidates. As precision medicine advances in autoimmune disorders, such network-guided strategies will be critical to translating omics data into biological and clinical insight. Beyond RA, this framework can be readily extended to other autoimmune conditions to uncover hidden mechanisms and inform biomarker discovery. Experimental validation, such as CRISPR perturbation assays or longitudinal patient data analysis, will be critical to evaluate the diagnostic or therapeutic potential of the identified candidates.

Unlike previous RA transcriptomic studies that relied mainly on differential expression or WGCNA \cite{zhang2005,wgcna}, our percolation-guided network construction and novelty screening revealed hub genes not previously reported, underscoring the added value of community detection approaches.

These high-novelty hubs may also hold promise as candidate biomarkers, though further validation is required.

\bigskip
\bibliographystyle{splncs04}
\bibliography{refs}

\end{document}